%% file: paper_5_8.tex
\documentclass[journal]{IEEEtran}
\usepackage{epsfig,latexsym,amssymb,amsmath,graphics,color}
\usepackage{graphicx,subfigure}
\usepackage{float}
\usepackage{verbatim}
\usepackage{multicol}
\usepackage{amstext}
\usepackage{mathrsfs}
\usepackage[all]{xy}
\usepackage{cite}
\usepackage{setspace}
\usepackage{amsthm}
\usepackage{algorithm}
\usepackage{algorithmicx}
\usepackage{algpseudocode}
\usepackage{booktabs}

\input{pream_bm}

\title{A 1Mbps Real-time NLOS UV Scattering Communication System with Receiver Diversity over 1km}
\author{Guanchu Wang, Kun Wang, Chen Gong, Difan Zou, Zhimeng Jiang and Zhengyuan Xu 
\thanks{This work was supported by National Key Basic Research Program of China
(Grant No. 2013CB329201), Key Program of National Natural Science Foundation
of China (Grant No. 61631018), Key Research Program of Frontier
Sciences of CAS (Grant No. QYZDY-SSW-JSC003), Key Project in Science
and Technology of Guangdong Province (Grant No. 2014B010119001), Shenzhen
Peacock Plan (No. 1108170036003286), and the Fundamental Research
Funds for the Central Universities.
The authors are with Key Laboratory of
Wireless-Optical Communications, Chinese Academy of Sciences, School of
Information Science and Technology, University of Science and Technology of
China, Hefei, China. Zhengyuan Xu is also with Shenzhen Graduate School,
Tsinghua University, Shenzhen, China. Email: hegsns@mail.ustc.edu.cn,
\{cgong821, xuzy\}@ustc.edu.cn.}}

\begin{document}
\begin{spacing}{1.0}

\maketitle

\begin{abstract}
In the non-line of sight (NLOS) ultraviolet (UV) scattering communication, the received signals exhibit the characteristics of discrete photoelectrons due to the extremely large path loss.
We design and demonstrate an NLOS UV scattering communication system in this work, where the receiver-side signal detection is designed based on a discrete-time Poisson channel model.
In our system, a laser and multiple photomultiplier tubes are employed as the optical transmitter and detector, respectively.
Furthermore, we design algorithms for pulse-counting, synchronization, channel estimation and $LLR$ computation for hardware realization in FPGA board.
Simulation results are provided to evaluate the proposed system design and specify the system key parameters.
We perform field tests for real-time communication with the transmission range over $1$km, where the system throughput reaches $1$Mbps.
\end{abstract}


\section{Introductions}

Non-line of sight (NLOS) ultraviolet (UV) scattering communication can maintain certain transmission rate for optical wireless communication (OWC) while not requiring perfect alignment of the transmitter and the receiver \cite{GA{2006}, ZXu{2008}}.
Prospective application of NLOS UV scattering communication includes the scenarios where radio-silence is required and the transmitter-receiver alignment is hard to guarantee due to obstacles or mobility.
Since most solar radiation in the spectrum between $200$nm and $280$nm is absorbed by the atmosphere \cite{ZXu{2008}}, the background solar radiation in such spectrum is typically negligible.
Thus NLOS UV scattering communication is competent for outdoor OWC under non-perfect transmitter-receiver alignment and strong solar background radiation.

NLOS UV scattering communication has been extensively investigated from both theoretic and experimental perspectives recently, which can be characterized by Poisson channel.
The capacities of continuous-time and discrete-time Poisson channel have been studied in \cite{Michael{2007}, Aaron{1988}} and \cite{JCao{2014}, ALapidoth{2009}}, respectively.
Moreover, the capacities of MISO and MIMO communication have been investigated in
\cite{SM{2003}} and \cite{KChakraborty{2008}}, respectively.
Regarding the channel characterization, the channel link gain and impulse response model have been extensively investigated in \cite{HDing{2009}, HZhang{2012}, YSun{2016}, RDrost{2013}, HXiao{2011}, YZuo{2013}}.
The receiver-side signal characterization and performance analysis has been studied in \cite{QHe{2010}}.
The generalized maximum-likelihood sequence detection has been discussed in \cite{NChatzidiamantis{2010}}, and the signal detection with receiver diversity has been investigated in \cite{Gong{2015}, ME{2015}}.
Besides the theoretical investigation, experiment works on the NLOS UV scattering channel characterization include the turbulence channel investigation \cite{LLiao{2015}, Kun{2017}}, the UV photon-counting based detection \cite{GA{2009}},  and long-distance channel characterization \cite{GChen{2014}}.
Research works on the experimental communication system realization to improve the transmission rate and the transmission range have been conducted since year $2000$ with several representative endeavors.
Experimental communication systems with transmission data rate of $2$kbps over $1$km distance has been reported in \cite{AM{2004}}.
Semi-real-time communication systems with nearly $80$kbps air-face data rate over $30$m NLOS transmission links, real-time experiments with nearly $2.4$kbps air-face data rate over $90$m LOS and $500$kbps air-face data rate over $50$m NLOS transmission links based on RS and LDPC code have been accomplished in \cite{DH2012, MW2014, LG2015}.
The link budget analysis based on experimental data for LDPC codes with air-face symbol rate $2$Mbps and distance over $80$m is performed in \cite{HQin{2017}}.
Furthermore, demonstration of $400$kbps throughput over transmission distance $500$m based on concatenated code has been realized in \cite{KunW{2017}}.

In this work, beyond the work \cite{KunW{2017}}, we extend the transmission distance and data rate of NLOS UV scattering communication to $1$km and $1$Mbps, respectively.
We still employ an UV laser as the transmitter, with an external-modulator for on-off keying (OOK) modulation.
Receiver diversity is adopted with equal gain combining since the signal intensities at different detectors are approximately the same.
We design and realize signal processing algorithms based on the discrete-time Poisson channel model.
Computer-based simulations are conducted to evaluate the performance of the designed communication system.
Furthermore, we realize a real-time point-to-point SIMO UV communication system consisting of multiple photomultiplier tubes (PMTs) with optical devices and amplifying circuits,
FPGA boards for implementation of digital signal processing,
and computers for real-time encoding and decoding of channel codes in software.
We complete a real-time communication experiment with the transmission distance over $1$km and the resulting throughput can reach $1$Mbps.

This paper is organized as follows.
In Section \uppercase\expandafter{\romannumeral2}, we provide the channel model and block diagrams of the proposed NLOS UV scattering communication system.
In Section \uppercase\expandafter{\romannumeral3}, we address the signal processing at the receiver side.
Simulation and real experimental measurement results are shown in Section \uppercase\expandafter{\romannumeral4}.
Finally, we conclude this paper in Section \uppercase\expandafter{\romannumeral5}.

\section{Signal Characterization and System Block Diagram}

\subsection{Discrete-time Poisson Channel with Receiver Diversity}

Consider the NLOS UV scattering communication system with receiver diversity.
We employ a UV laser with constant power $\widetilde{P}_t$ at the transmitter side.
An external modulator is used for the on-off keying (OOK) modulation, which changes the direction of the UV light beam for symbol $s_k$, such that for $s_k = 0$ the light can be blocked at the transmitter side while for $s_k = 1$ the light can be emitted.
At the receiver side, multiple detectors are employed to increase the received signal intensity.
Due to the extremely large path loss of the NLOS scattering communication, the detected signal can be characterized by discrete photoelectrons whose number satisfies a Poisson distribution.
We assume that given the transmitted symbols and the link gains of the receivers, the numbers of photoelectrons for different detectors are independent of each other due to independent photon reception and photoelectric conversion of different detectors.

Assume that $K$ detectors are employed at the receiver side.
Let $N_k$ denote the number of received photoelectrons from detector $k$, which satisfies the following Poisson distribution
\begin{equation}
\begin{aligned}
\label{eq:Poisson_channel}
\mathbb{P} (N_k = n | s = 1) &= \frac{(\lambda_{s,k} + \lambda_{b,k})^n}{n!} e^{-(\lambda_{s,k} + \lambda_{b,k})},
\\
\mathbb{P} (N_k = n | s = 0) &= \frac{\lambda_{b,k}^n}{n!} e^{-\lambda_{b,k}},
\end{aligned}
\end{equation}
where $\lambda_{s,k}$ and $\lambda_{b,k}$ denote the mean numbers of detected photoelectrons at detector $k$ for the signal and background radiation components, respectively.
For the transmission symbol rate $R_s$ under OOK modulation, we have that
\be
\lambda_{s,k} = \frac{\eta_k \widetilde{P}_t}{\xi_k h \nu R_s},
\ee
where $\eta_k$, $h$, $\nu$ and $\xi_k$ denote the quantum efficiency of detector $k$ including optical filter and photon-detector, Planck's constant, the frequency of the optical signal and the path loss from the transmitter to detector $k$, respectively.

\subsection{Communication System Block Diagram}

At the transmitter side, the information bits are randomly generated and encoded in the transmitter-side personal computer (TX-PC).
The encoded blocks are divided into transmission frames and sent from the TX-PC to the FPGA board via ethernet cable.
On the FPGA board, the synchronization and pilot sequences are added at the head of each frame and certain protection intervals are added between consecutive frames.
The frames with protection intervals are exported via the pin of FPGA board, which drive the external modulator of the UV laser.

\begin{figure*}
\centering
	\includegraphics[width=1.0\textwidth]{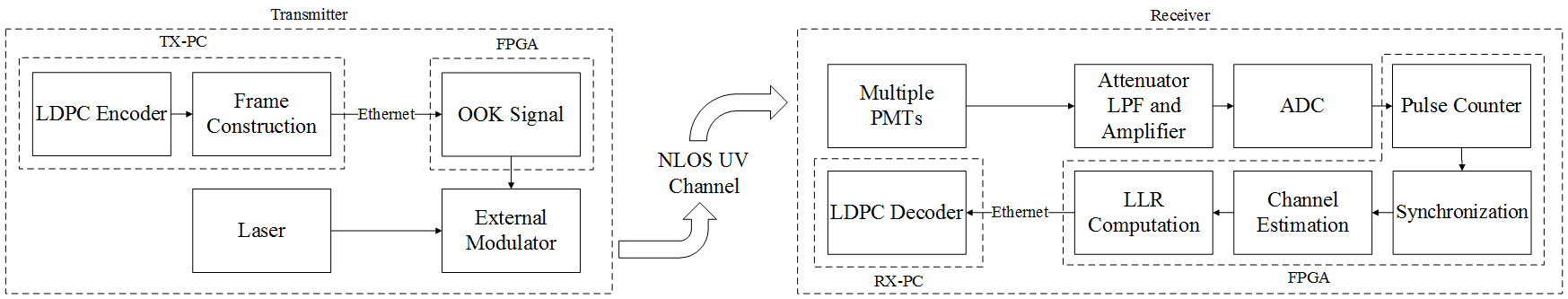}
    \caption{\label{fig:transmitter_receiver_structure}The block diagram of NLOS UV scattering communication system.}
\end{figure*}

\begin{figure*}
\centering
	\includegraphics[width=1.0\textwidth]{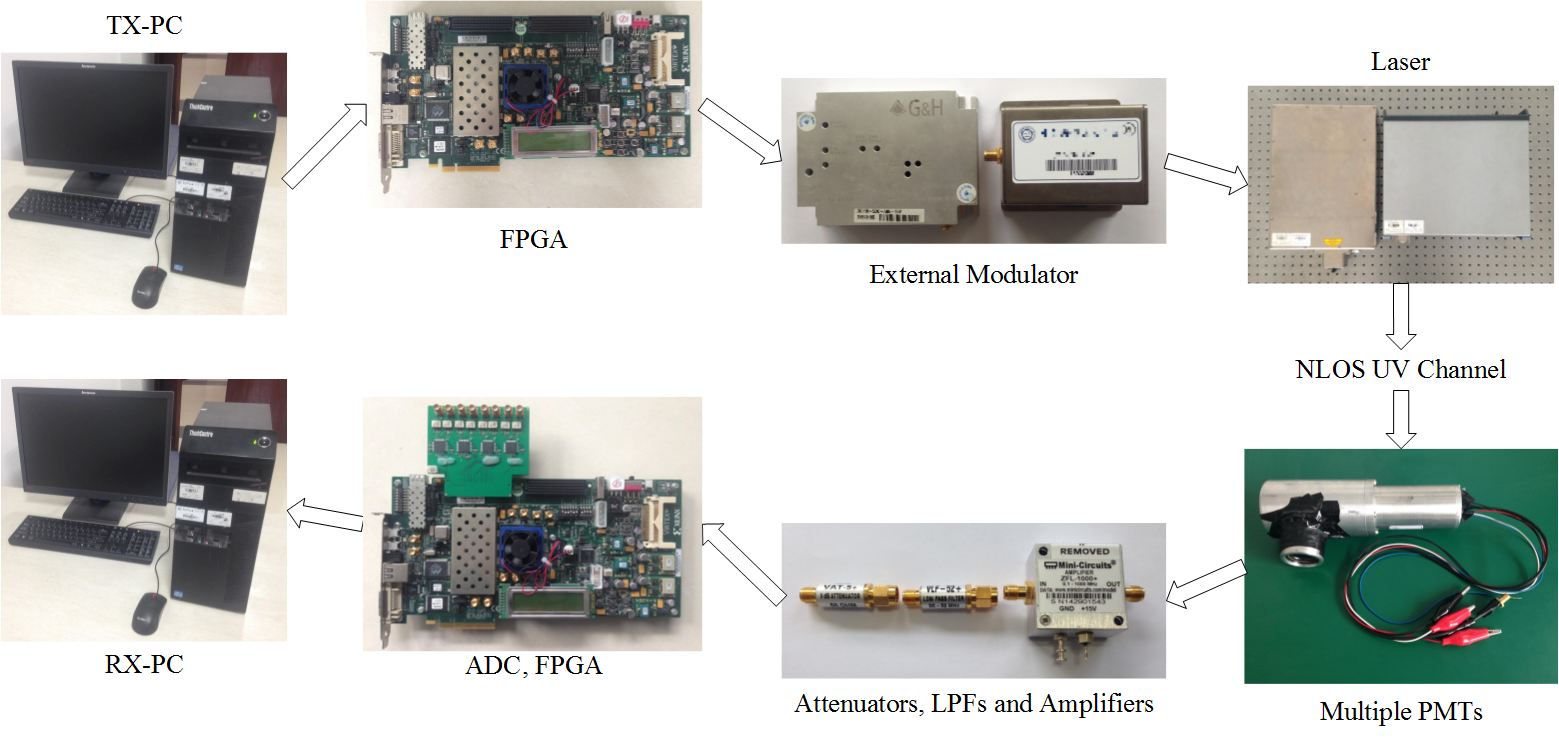}
    \caption{\label{fig:total_system}The hardware realization blocks for the NLOS UV scattering communication system.}
\end{figure*}

At the receiver side, PMTs are employed as the photon-detector, which can convert the received photoelectrons to analog pulses such that photon counting can be realized via pulse counting.
A HAMAMATSU PMT (R7154 module) detector is integrated with an optical filter and a sealed box, which passes the light signal in the UV spectrum of wavelength around $266$nm and blocks the background radiation in other spectrum.
To enable effective pulse counting process, attenuators, low-pass filters and amplifiers are adopted.
The pulses are firstly attenuated by the electrical attenuator such that it falls into the input range of the amplifier and then amplified.
The amplified pulses pass through the low-pass filter such that the pulse width can be increased and the rising edge detection for pulse counting can be performed more accurately.
Analog-to-digital converters (ADCs) are employed such that the rising edge detection can be performed in the digital domain.
The digital processing of synchronization, channel estimation and symbol detection are performed sequentially based on discrete-time Poisson channel characterization in the FPGA board.
To achieve the processing for receiver diversity, a Xilinx Virtex-6 ML605 board is adopted to support multiple ADCs.
The soft information of received signal detection is packed at the FPGA board and sent to the receiver-side personal computer (RX-PC) through the ethernet cable, which is employed for decoding at the receiver side.
The block diagram and experimental system of the entire NLOS UV scattering communication link are shown in Figure \ref{fig:transmitter_receiver_structure} and Figure \ref{fig:total_system}, respectively.

\section{Digital Signal Processing in Hardware Realization}

We outline the digital signal processing in FPGA hardware realization, including the pulse counting with receiver diversity combining, the counting-based synchronization, the counting-based channel estimation and LLR computation, and the LDPC construction and decoding.

\subsection{Pulse-count and Receiver Diversity Combining}

Based on the output waveform of the AD-convertor for each PMT, the pulse counting can be realized according to the rising-edge detection with a voltage threshold $V_{thd}$.
A pulse can be recorded if the previous sample voltage is lower than $V_{thd}$ and the current sample is higher than $V_{thd}$.

Note that since the $K$ PMTs are placed close to each other, the received signal intensities are virtually the same.
Then the maximum-likelihood detection among the $K$ detectors can be approximated by equal gain combining, which is adopted for $K$ PMTs in this work.

Let $N_{k}$ denote the number of pulses of PMT $k$ in a symbol duration.
For the equal gain combining, letting $N = \sum^K_{k=1} N_{k}$, with the signal component intensity $\lambda_s \triangleq \sum^K_{k=1} \lambda_{s,k}$ and noise component intensity $\lambda_b \triangleq \sum^K_{k=1} \lambda_{b,k}$, we have that $N$ satisfies a Poisson distribution with mean $(\lambda_s + \lambda_b)$ for OOK symbol on and the mean $\lambda_b$ for OOK symbol off.
Due to the Poisson channel characteristics, such signal combination can significantly improve the communication system performance.

\subsection{Counting-based Synchronization}

Note that the pulse-counting has been investigated in \cite{XiaonaLiu{2016}}. We employ pulse-counting-based digital signal processing for synchronization.
Since the starting time of a frame is difficult to detect, in this work the received signal is considered to be divided into chips with duration $T_c = \frac{T_s}{M}$, where $T_s$ denotes the duration of one symbol.
For certain chip index $t$, the numbers of pulses in the past $LM$ consecutive chips are reserved in an $L \times M$ matrix $\bC_t$, where each element denotes the number of pulses in chip $t - (i-1)M - j + 1$ for $1 \leq i \leq L$ and $1 \leq j \leq M$.
Let $\bs$ denote the chip-level synchronization sequence containing the $LM$ consecutive chips corresponding to the $L$ binary symbols.
The synchronization is formulated by the following maximum auto-correlation problem,
\be
\label{eq:synchronization_theroy}
\hat{t} = \arg \max_{t} \big\{ (2 \bs^T - \boldsymbol{1}^T_L) \bC_t \boldsymbol{1}_M \big\},
\ee
where $\boldsymbol{1}^T_L$ and $\boldsymbol{1}_M$ are all-one $L$-dimensional row and $M$-dimensional  column vector, respectively; and $\hat{t}$ denotes the estimated ending index of the synchronization sequence.

The synchronization based on Equation (\ref{eq:synchronization_theroy}) requires finding the optimal $\hat{t}$ with the maximum correlation peak among all possible positions, which leads to a large delay in hardware realization.
To reduce the delay, we adopt reduced-complexity synchronization based on shortening the search window.
Let $C_{thd}$ denote a threshold for activating the peak value searching, and $W$ denote the width of the searching window, which are pre-determined via training from existing experimental results.
The search on the maximum correlation peak is activated in the interval $[\tilde{t}, \tilde{t} + WT_c)$, once $\bs^T \bC_{\tilde{t}} \boldsymbol{1}_M > C_{thd}$.
The performance of proposed algorithm depends on the preset values of $C_{thd}$ and $W$.
The time delay $WT_c$ is brought by the proposed peak value searching.

\subsection{Channel Estimation and LLR Coumputation}

The synchronization sequence is also adopted for channel parameter estimation.
Letting $\bC_{\boldsymbol{s}}$ denote the number of pulses in the pilot symbols corresponding to synchronization sequence $\bs$, we have $\bC_s = \bC_{\hat{t} - WT_c}$ due to the synchronization delay, where the number of pulses for each symbol can be obtained via merging those of consecutive $M$ chips.
Hence the mean of signal component $\lambda_s$ and background radiation $\lambda_b$ for each symbol can be estimated by the following unbiased estimation,
\begin{equation}
\begin{aligned}
\label{eq:channel_estimation_theory}
\widehat{\lambda}_b + \widehat{\lambda}_s &= \frac{\bs^T \bC_{\boldsymbol{s}} \boldsymbol{1}_M}{\bs^T \boldsymbol{1}_L},
\\
\widehat{\lambda}_b &= \frac{\bar{\bs}^T \bC_{\boldsymbol{s}} \boldsymbol{1}_M}{\bar{\bs}^T \boldsymbol{1}_L},
\end{aligned}
\end{equation}
where $\bar{\bs} = \boldsymbol{1}_L - \bs$.
Since $N$ has been denoted as the number of pulses in one symbol, according to Equation (\ref{eq:Poisson_channel}), the log-likelihood ratio ($LLR$), denoted as $\mathcal{L} (N)$, is given by
\be
\label{eq:llr_cnt}
\mathcal{L} (N) = N \log \big( \frac{\widehat{\lambda}_s + \widehat{\lambda}_b}{\widehat{\lambda}_b} \big) - \widehat{\lambda}_s
\sim N - \frac{\widehat{\lambda}_s}{\log \big( \frac{\widehat{\lambda}_s + \widehat{\lambda}_b}{\widehat{\lambda}_b} \big)}.
\ee

The hardware realization of channel estimation and $LLR$-computation is based on Equations (\ref{eq:channel_estimation_theory}) and (\ref{eq:llr_cnt}), respectively. Since the division operation is difficult to realize in a real-time high-speed manner, to reduce the computational complexity of channel estimation, the quantized results of division $\theta^s_i = (p_s i)^{-1}$ and $\theta^b_i = (p_b i)^{-1}$ for $1 \leq i \leq L$ are computed in an offline manner and stored in a table, where $p_s$ and $p_b$ are the preset precision on the estimation of $\lambda_s$ and $\lambda_b$, respectively.
The estimation for channel parameters is given by $\widehat{\lambda}_s = \theta^s_{\boldsymbol{s}^T \boldsymbol{1}_L} \bs^T \bC_{\boldsymbol{s}} \boldsymbol{1}_M$ and $\widehat{\lambda}_b = \theta^b_{\bar{\boldsymbol{s}}^T \boldsymbol{1}_L} \bar{\bs}^T \bC_{\boldsymbol{s}} \boldsymbol{1}_M$, respectively.

For the low-complexity realization of $LLR$-computation, let $\phi_{i,j} = \lceil \frac{p_s i}{\log(\frac{p_s i + p_b j}{p_b j})} \rceil$, for $1 \leq i \leq \Lambda_s$ and $1 \leq j \leq \Lambda_b$, where $\Lambda_s$ and $\Lambda_b$ denote the preset maximum estimation value of $\lambda_s$ and $\lambda_b$, respectively.
The values of $\phi_{i,j}$ are computed offline and stored in a table.
We then compute $LLR$ via $\mathcal{L} (N) = N - \phi_{\lfloor \widehat{\lambda}_s \rfloor, \lfloor \widehat{\lambda}_b \rfloor}$ and store the results into a buffer.
Once having collected the $LLR$s corresponding to one LDPC coded block, the messages are packed and sent to the RX-PC for LDPC decoding via the ethernet cable from the FPGA board.

\subsection{LDPC Code Construction and Decoding}

A $(n_c, k_c)$ irregular LDPC code is employed for error correction, where $n_c$ and $k_c$ denote the number of coded and information symbols, respectively.
We consider the construction of parity-check matrix proposed in \cite{MY{2004}}.
The proposed construction of the parity check matrix $\bH_{(n_c-k_c)\times n_c} = [\bH_1, \bH_2]$, where $\bH_1$ is a regular block-wise Quasi-Cyclic (QC) low-density circular matrix \cite{MPCF{2004}}.
A $(J, D)$-regular QC parity-check matrix $\bH_1$ is specified as follows,
\be
\bH_1 =
\left[ \begin{array}{cccc}
\bI_p(0)	       & \bI_p(0)           & \ldots        & \bI_p(0)\\
\bI_p(0)  		   & \bI_p(p_{1,1})     & \ldots        & \bI_p(p_{1,D-1}) \\
\vdots             & \vdots             & \ddots        & \vdots\\
\bI_p(0)           & \bI_p(p_{J-1,1})   & \ldots        & \bI_p(p_{J-1,D-1}) \\
\end{array} \right],
\ee
where $\bI_p(p_{j,l})$ represents the circulant permutation matrix with element one at column $[(r + p_{j,l}) \mod p]$ in row $r$, for $1 \leq r \leq p-1$, $1 \leq j \leq J-1$ and $1 \leq l \leq D-1$; and $\bI_p(0)$ denotes a $p \times p$ identity matrix.
For the girth larger than or equal to $6$, we set $p_{j, l} = [(q^j_1 - 1)(q^l_2 - 1) \mod p]$ for $1 \leq j \leq J-1$ and $1 \leq l \leq D-1$, where $p$ is a prime number and $q_1$ and $q_2$ are two nonzero distinct elements of $\textbf{GF} (p)$.
Moreover, $\bH_2$ is a $(n_c - k_c) \times (n_c - k_c)$ square sawtooth matrix given by
\be
\bH_2 =
\left[ \begin{array}{cccccccc}
&1\\
&1     &1\\
&      &1    &1\\
&      &     &     &\ldots\\
&      &     &     &     &1    &1\\
&      &     &     &     &     &1    &1\\
\end{array} \right].
\ee
The systematic generator matrix $\bG_{k_c \times n_c}$ corresponding to above parity check matrix $\bH$ is given by $\bG_{k_c \times n_c} = [\bI, \bH^T_1 \bH^{-T}_2]$, where $\bH^{-T}_2$ is the binary transposed inverse of $\bH_2$ given by
\be
\bH^{-T}_2 =
\left[ \begin{array}{cccccc}
&1      &1     &\ldots &\ldots &1\\
&       &1     &\ldots &\ldots &1\\
&       &      &       &\ldots & \\
&       &      &       &1      &1 \\
&       &      &       &       &1 \\
\end{array} \right].
\ee

We adopt the reduced-complexity message passing algorithm (MPA) for the LDPC decoding proposed in \cite{JC{2005}}.
For the Tanner graph corresponding to the parity-check matrix $\bH$, let $\mathcal{M}(n)$ and $\mathcal{N}(m)$ denote the set of check nodes linked to the $n^{th}$ variable node and the set of variable nodes connected to the $m^{th}$ check node.
Let $L_{m \to n} (N_n)$ and $Z_{n \to m} (N_n)$ denote the message passed from the $m^{th}$ check node to the $n^{th}$ variable node and from the latter to the former, respectively.

The updating of passing message for the MPA is shown as follows.
Initialize $L_{m \to n} (N_n) = 0$ and $Z_{n \to m} (N_n) = \mathcal{L} (N_n)$.
The external message update at the $n^{th}$ check node for $n \in \mathcal{N}(m)$ is given in the Equation (\ref{eq:code_L_m2n}),
\begin{figure*}
\be
\label{eq:code_L_m2n}
L_{m \to n} (N_n) = \bigg( \prod_{n' \in \mathcal{N}(m) \backslash n} \sgn \big( Z_{n' \to m} (N_{n'}) \big)  \bigg) \times \frac{\min\limits_{n' \in \mathcal{N}(m) \backslash n} | Z_{n' \to m} (N_{n'}) |}{\alpha}
\ee
\end{figure*}
where $\alpha$ is a normalization constant larger than one.
The external message update at the $m^{th}$ variable node for $m \in \mathcal{M}(n)$ is given by
\be
\label{eq:code_Z_n2m}
Z_{n \to m} (N_n) = \mathcal{L} (N_n) + \sum_{m' \in \mathcal{M}(n) \backslash m} L_{m' \to n} (N_n).
\ee
The decision for each symbol is based on the following equation,
\be
\label{eq:code_Z_n}
Z_n (N_n) = \mathcal{L} (N_n) + \sum_{m \in \mathcal{M}(n)} L_{m \to n} (N_n).
\ee
The detection result $\hat{\bs} \triangleq [\hat{s}_1, \hat{s}_2, \ldots, \hat{s}_{n_c}]^T$, where $\hat{s}_n = \sgn \big[ Z_n (N_n) \big]$ for $1 \leq n \leq n_c$.
The symbol detection is performed in each iteration, and the MPA terminates if the parity check relation is satisfied, i.e.,
$\bH \hat{\bs} = \boldsymbol{0}$.
The decoding failure is recorded if the parity check relation has not been satisfied within a preset threshold of iteration times, which is typically set to be $10$ or $20$ in practical implementation.

\begin{figure}[H]
\centering
\begin{minipage}{1.0\linewidth}
	\includegraphics[width=1.0\textwidth]{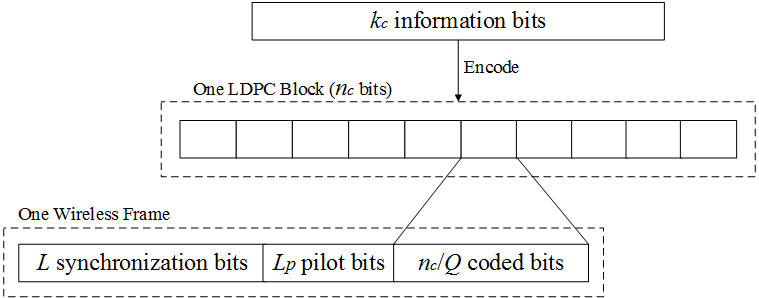}
    \caption{\label{fig:wireless_frame_structure}The structure of a transmission frame.}
\end{minipage}
\begin{minipage}{1.0\linewidth}
	\includegraphics[width=1.0\textwidth]{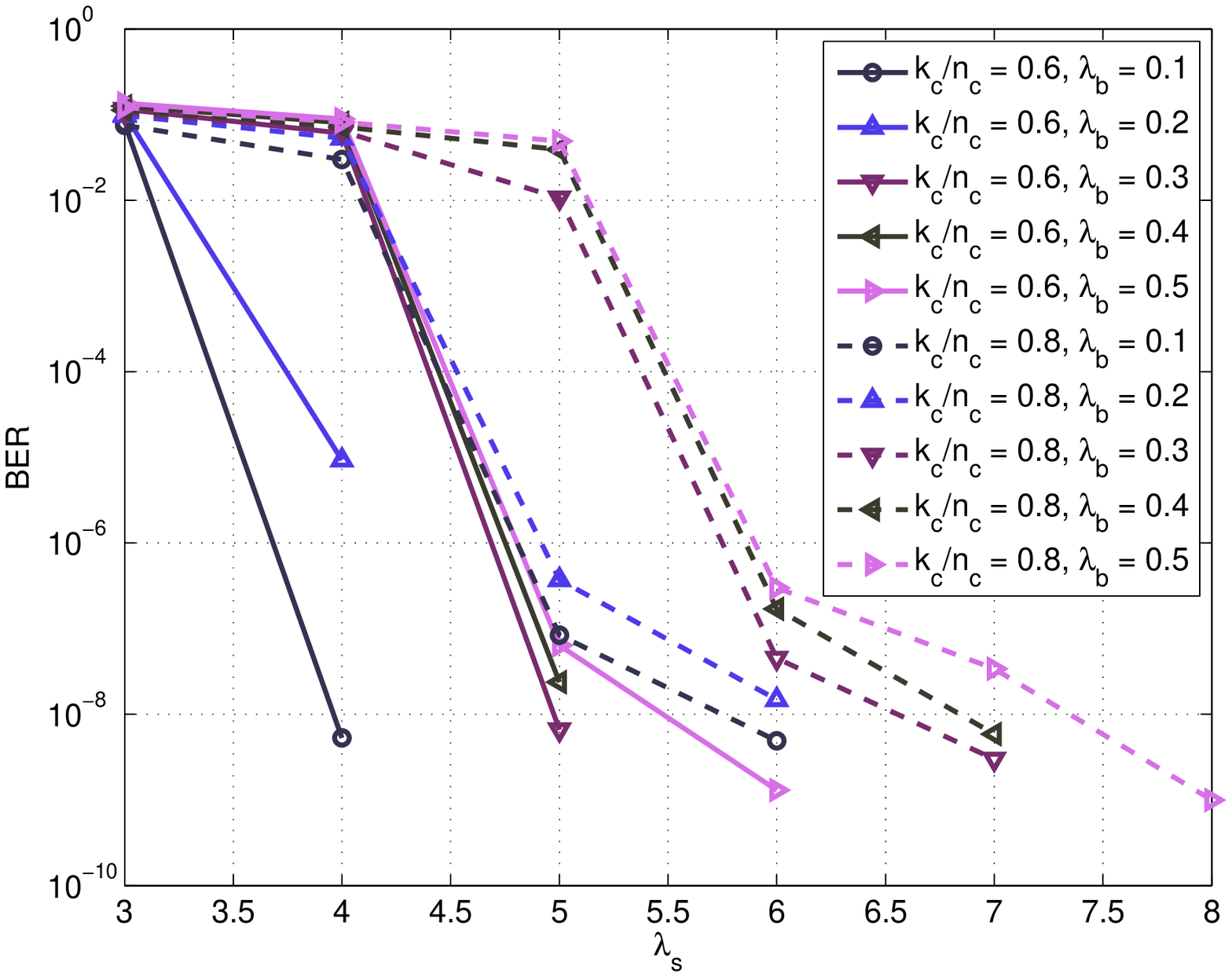}
    \caption{\label{fig:LDPC_Poisson}The BER of LDPC code on Poisson channel for different code rates.}
\end{minipage}
\end{figure}

Since the LDPC block length typically reaches several thousands to ten thousands, each block needs to be divided into $Q$ segments for transmission.
Synchronization and indication sequences are then added in front of each segment, where the indication sequence stores the block location of each frame in the LDPC block.
Each frame contains $(\frac{n_c}{Q} + L + L_p)$ bits, where $L$ and $L_p$ are the lengths of synchronization sequence and indication sequence, respectively.
The structure of a frame is shown in Figure \ref{fig:wireless_frame_structure}.

\section{System Design with Simulation and Experimental Results}

\subsection{System Specification with Simulation Results}

\begin{figure}
\centering
\begin{minipage}{1.0\linewidth}
	\includegraphics[width=1.0\textwidth]{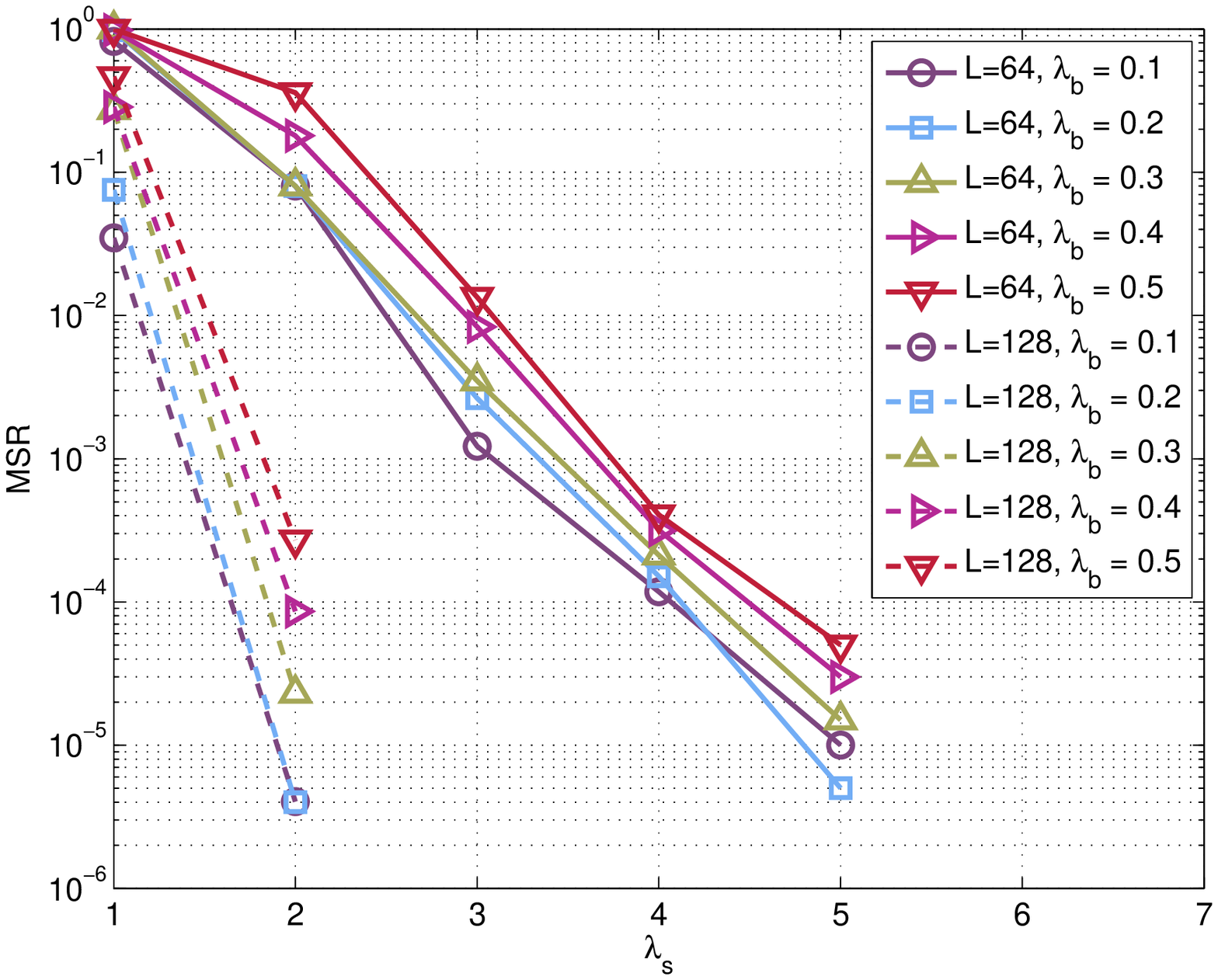}
    \caption{\label{fig:simulation_missframe}The miss synchronization rata versus $\lambda_s$ for different $\lambda_b$.}
\end{minipage}
\begin{minipage}{1.0\linewidth}
	\includegraphics[width=1.0\textwidth]{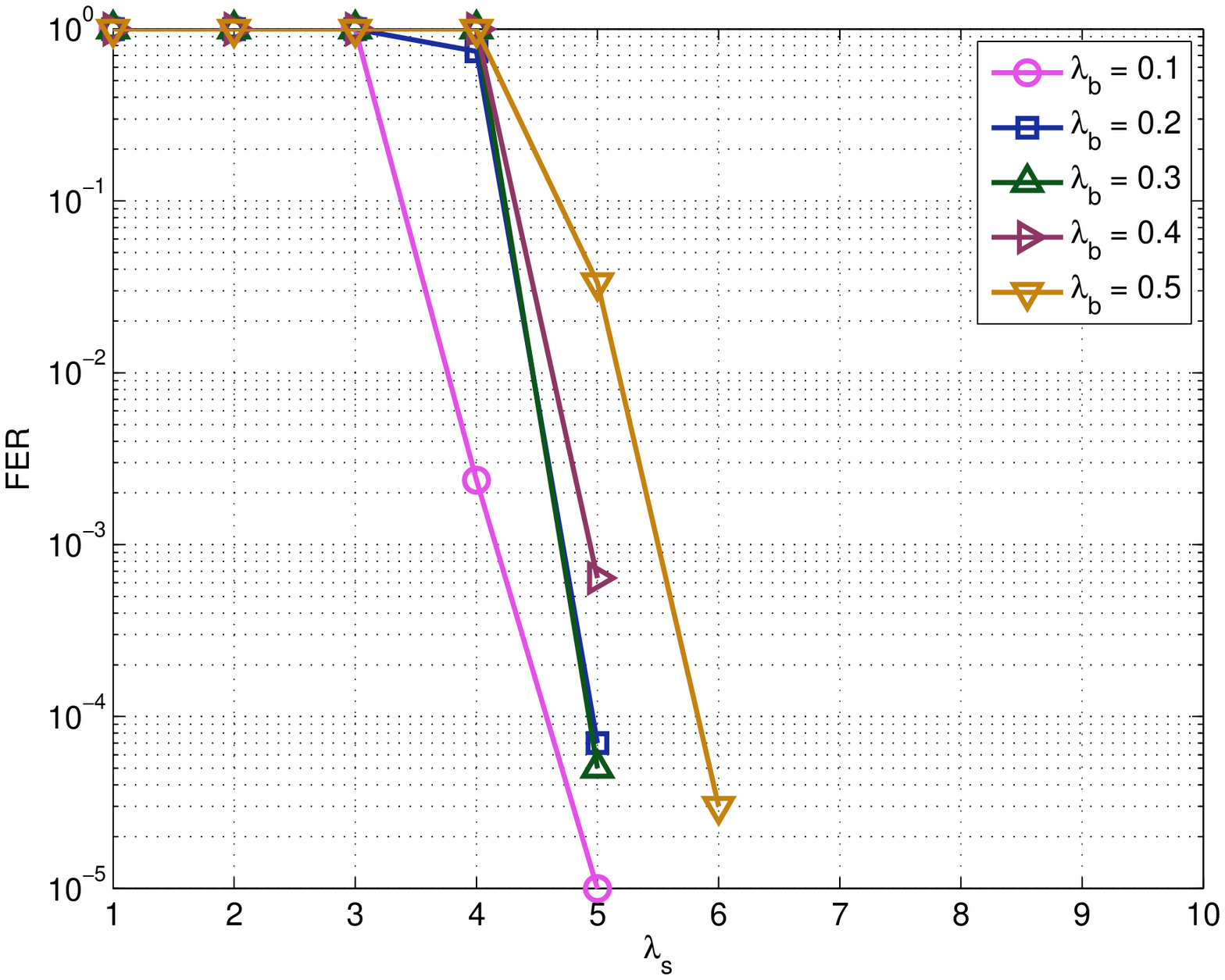}
    \caption{\label{fig:simulation_fer}The FER against $\lambda_s$ for different $\lambda_b$.}
\end{minipage}
\end{figure}

We specify the NLOS UV communication system under consideration.
Let the number of detectors $K = 3$, the transmission symbol rate of OOK modulation $R_b = 2M$bps, and the number of chips within a symbol duration $M = 10$.
Note that the solar background radiation is approximately $1 \times 10^4$ to $5 \times 10^4$ counts per second at the wavelength $266$nm.
Due to the inter-symbol interference from the low-pass properties of the external modulator, around half chip of pulses for OOK symbol off is affected by the previous symbol on, which implies that the background radiation intensity can increase via an amount $\frac{\lambda_s}{2M}$ if the inter-symbol interference is not taken into consideration.
Hence we adopt certain margin for the background radiation intensity within each symbol duration, and set $\lambda_b \in \{ 0.1, 0.2, 0.3, 0.4, 0.5 \}$.
For rate-$0.6$ $(12630, 7578)$ and rate-$0.8$ $(12630, 10104)$ LDPC codes, we show the BER versus $\lambda_s$ in Figure \ref{fig:LDPC_Poisson}; and for the length of synchronization sequence $L = 64$ and $L = 128$, we plot the miss synchronization rate (MSR) against $\lambda_s$ in Figure \ref{fig:simulation_missframe}.
It is observed that for rate-$0.6$ code, the BER can drop below $10^{-6}$ when $\lambda_s \ge 5.0$, where $L = 64$ suffices to suppress the MSR below $10^{-4}$.

Based on above simulation results, we adopt the length of synchronization sequence $L = 64$, and the LDPC code with parameters $J = 12$, $D = 18$, $p = 421$, $k_c = 7578$, $n_c = 12630$ and code rate $k_c/n_c = 0.6$.
More concrete realization parameters include: the precision of channel estimation $p_s = 0.5$, $p_b = 0.01$, the segment number $Q = 10$ and the pilot length $L_p = 17$.
For such coding scheme, the simulation results for the frame error rate (FER) is shown in Figure \ref{fig:simulation_fer}.

\subsection{Lab Test on the Communication System Performance}

We conduct real-time communication performance lab test on the system performance before outdoor field test, with the designed system parameters from the simulation results.
The entire experimental system is shown in Figure \ref{fig:total_system}.
Limited by the laboratory space, two OD3 UV optical decay plates are employed to emulate the weak link gain of the scattering channel, where the attenuation factor for each one is $3.4 \times 10^3$ and thus the entire path loss is over $1.16 \times 10^7$.
The sampling rate of ADC module is $100$MHz, the wavelength of UV light is $266$nm, the transmission power is $30$mW, and the air-face transmission symbol rate is $2$Mbps.

\begin{figure}
\centering
\begin{minipage}{1.0\linewidth}
	\includegraphics[width=1.0\textwidth]{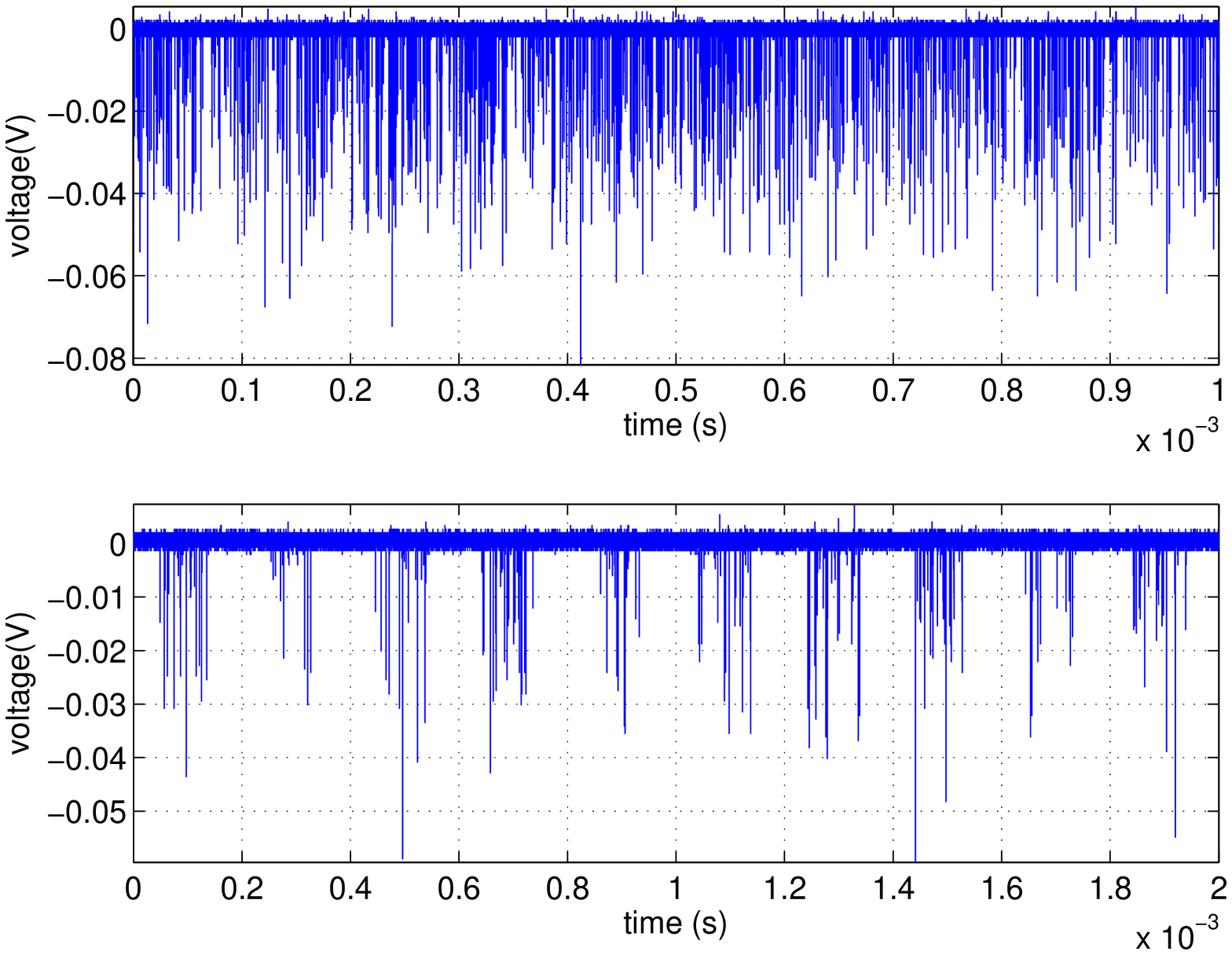}
    \caption{\label{fig:PMT_waveform}The output waveform of a PMT with detection of constant signal (top) and OOK-modulated signal (bottom).}
\end{minipage}
\begin{minipage}{1.0\linewidth}
	\includegraphics[width=1.0\textwidth]{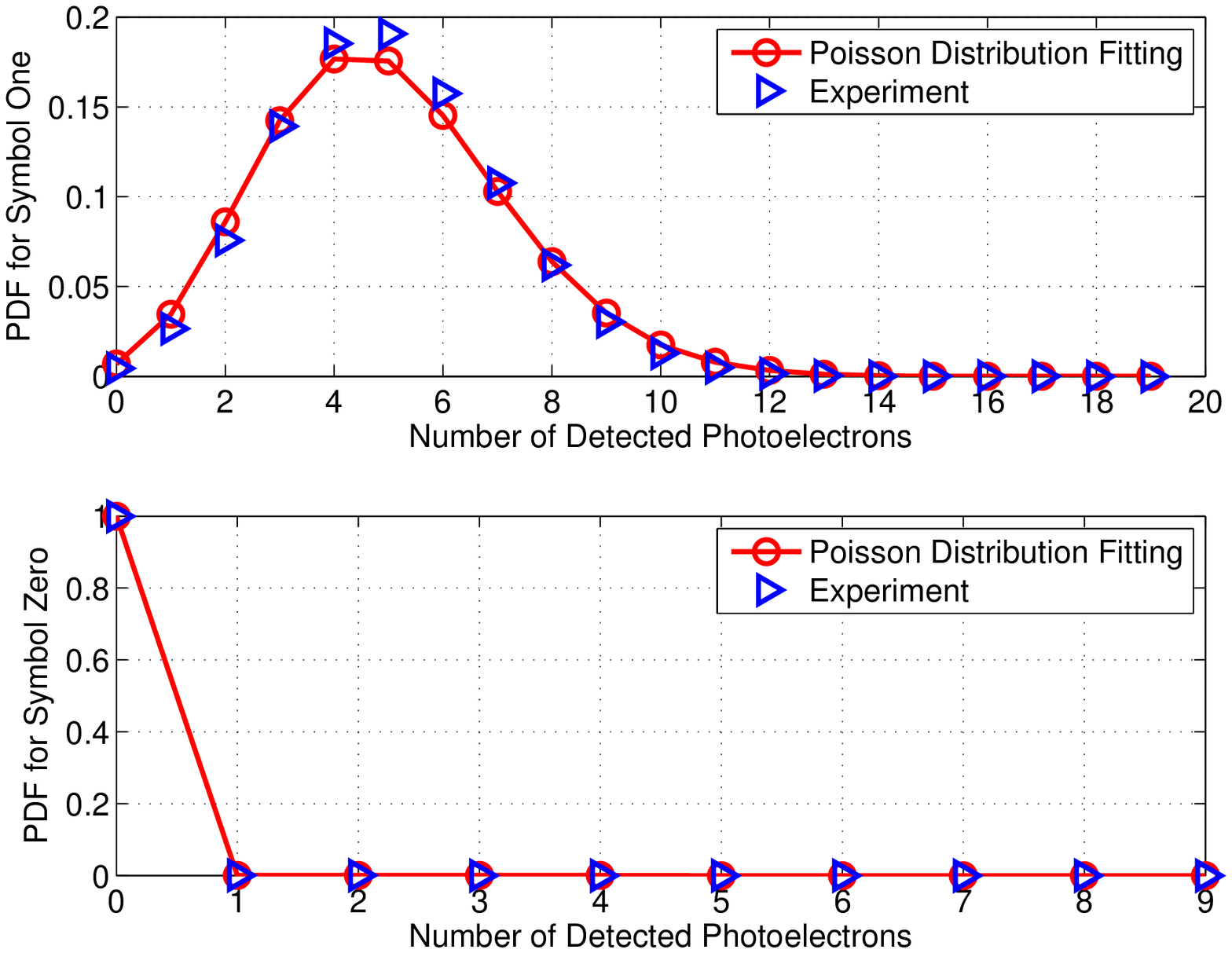}
    \caption{\label{fig:Poisson_OOK}The distribution on the number of detected photoelectrons for OOK Symbol One (top) and Zero (bottom).}
\end{minipage}
\end{figure}

Figure \ref{fig:PMT_waveform} plots the output waveform of a PMT under constant UV light power and OOK-modulated signals.
Each negative pulse corresponds to one or several detected photoelectrons, where the pulse duration is $10$ns.
The distributions on the number of pulses for OOK symbol on and background radiation are illustrated in Figure \ref{fig:Poisson_OOK}, where Poisson distribution fitting is also plotted for comparison.
We have also conducted real-time communication for lab test and obtained the following results.
For totally $2 \times 10^4$ transmitted frames, miss synchronization occurs for fewer than $50$ frames, and decoding error occurs for only $1$ frame.
Based on above test results, we conclude that the throughput is nearly $\big[ 1 - (50+1)/(2 \times 10^4) \big] \times 1263/(1263+17+64) \times 0.6 \times 2$Mbps $\approx 1.125$Mbps.

\subsection{Field Test over 1km}

\begin{figure}
\centering
\begin{minipage}{1.0\textwidth}
	\includegraphics[width=0.49\textwidth]{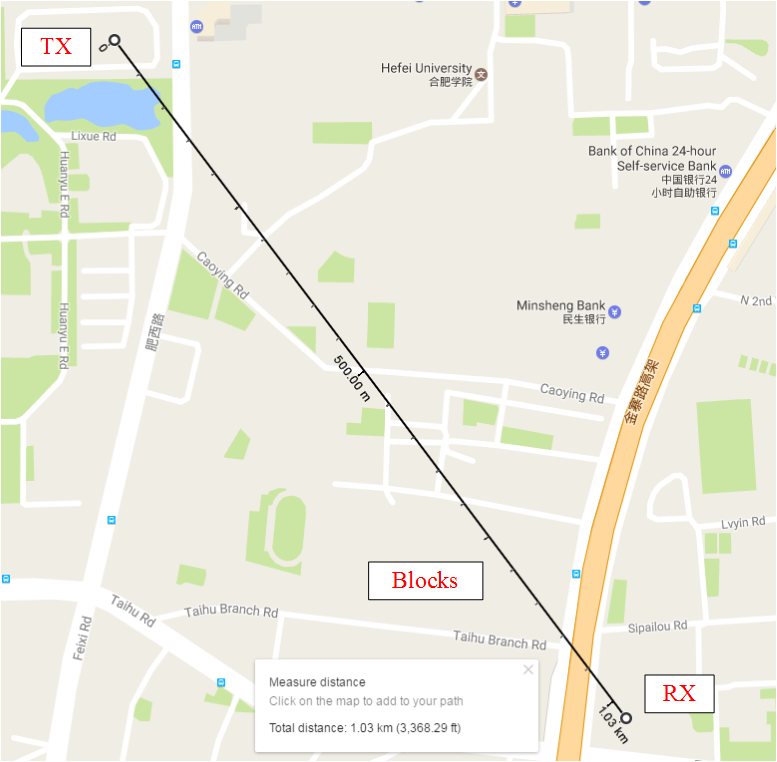}
    \caption{\label{fig:google_map}The location of transmitter and receiver.}
\end{minipage}
\begin{minipage}{1.0\textwidth}
	\includegraphics[width=0.49\textwidth]{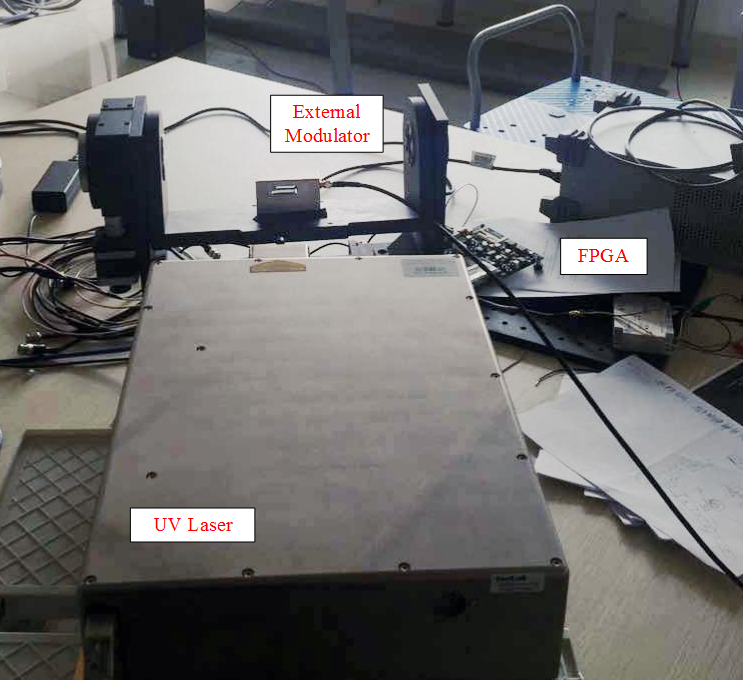}
    \caption{\label{fig:Tra}The transmitter-side test bed.}
\end{minipage}
\end{figure}

\begin{figure}
\centering
	\includegraphics[width=0.49\textwidth]{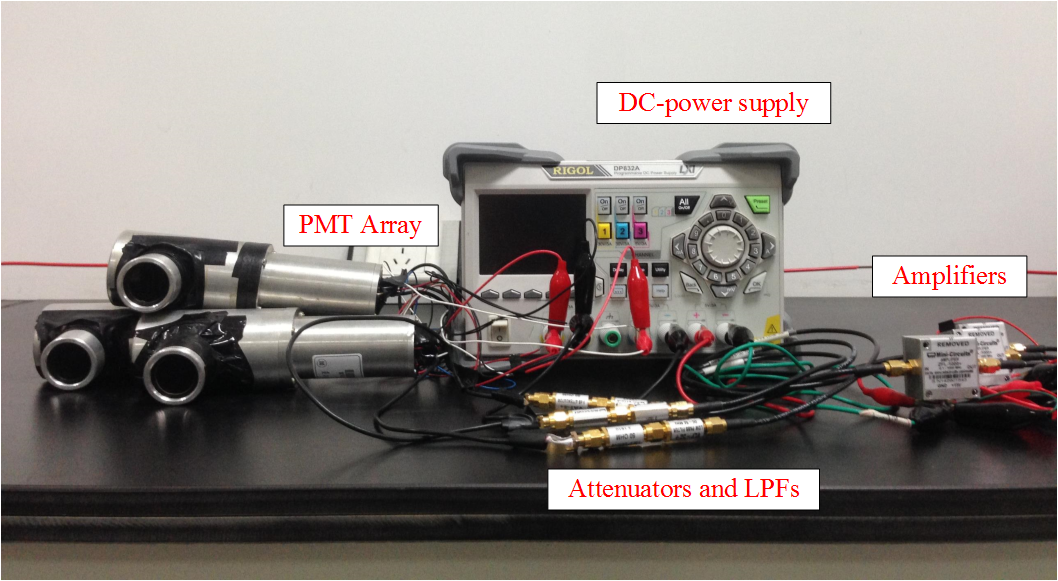}
	\includegraphics[width=0.49\textwidth]{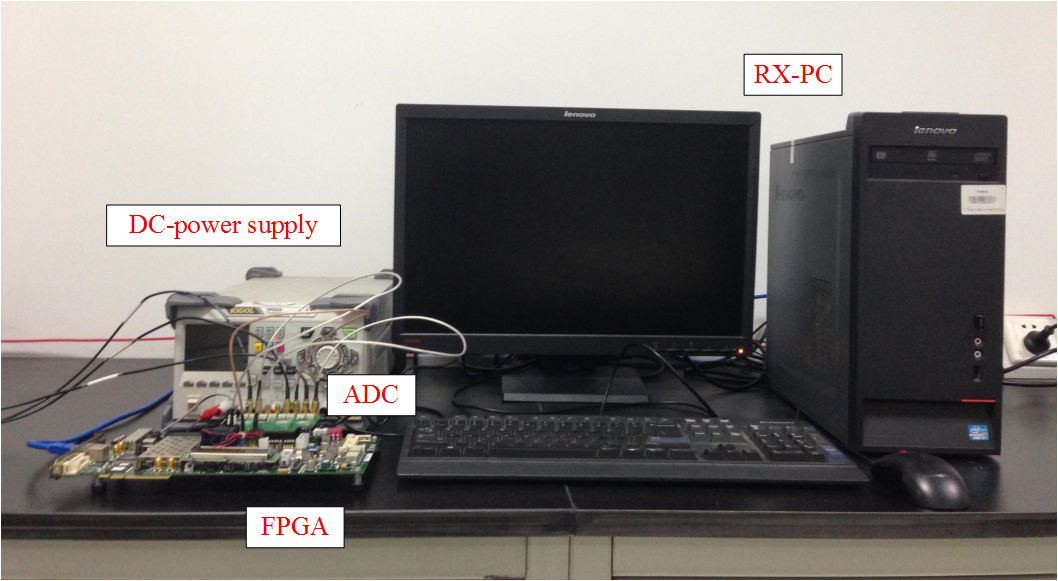}
\caption{\label{fig:Rcv2}The receiver-side test bed.}
\end{figure}

We carried out the outdoor field test on the real-time UV communication.
The locations of transmitter and receivers were placed in two buildings shown in Figure \ref{fig:google_map}.
The transmission distance was $1.03$ kilometers and about $1$ degree offset angle was adopted.
The date of outdoor field test was Apr. $21^{st}$ in $2017$, where the weather conditions were $11^{\circ}$C$-22^{\circ}$C, cloudy, and northwest wind of $5.5 \sim 7.9$m/s.
All parameters for outdoor field test are shown in Table \uppercase\expandafter{\romannumeral1}.
We show the transmitter-side and receiver-side test bed in Figures \ref{fig:Tra} and \ref{fig:Rcv2}, respectively.

\begin{table}[htbp]
\centering
\label{tab:test}
\begin{tabular}{cc}
\hline
   Parameters            & Value \\
\hline
   Transmission Power    & $120$mW \\
   Transmission Range    & $1.03$km \\
   Scattering Angle      & $\approx 1^{\circ}$ \\
   UV Light Wavelength   & $266$nm \\
   Air-face Symbol Rate  & $2$Mbps \\
   Total Number of Frames & $1.7 \times 10^4$ \\
   Number of Miss Synchronization & $221$ \\
   Number of Error Frames & $5$ \\
\hline
\end{tabular}
\caption{Parameters and Results of outdoor field test.}
\end{table}

The results of field test are shown in Table \uppercase\expandafter{\romannumeral1}.
For totally $1.7 \times 10^4$ transmitted frames, miss synchronization occurs for $221$ frames, and decoding error occurs for only $5$ frames.
Hence the results imply that the throughput is nearly $\big[ 1 - (221+5)/(1.7 \times 10^4) \big] \times 1263/(1263+17+64) \times 0.6 \times 2$Mbps $\approx 1.113$Mbps.

\section{Conclusions}

We have designed the system and finished hardware realization based on receiver diversity for the NLOS UV scattering communication over $1$km, where the system throughput can reach more than $1$Mbps.
Higher data rate and longer transmission range can be achieved if more powerful optical devices and more capacity approaching coded modulation are employed, which remains for future work.

\end{spacing}
\end{document}

%% file: pream_bm.tex
\newtheorem{prop}{Proposition}

\newtheorem{cor}{Corollary}

\newtheorem{lm}{Lemma}

\newtheorem{thm}{Theorem}

\newcommand{\be}{\begin{eqnarray}}
\newcommand{\ee}{\end{eqnarray}}
\newcommand{\benn}{\begin{eqnarray*}}
\newcommand{\eenn}{\end{eqnarray*}}
\def\IR{\rm I \kern-0.20em R}
\newcommand{\utwi}[1]{\mbox{\boldmath $ #1$}}

\newcommand{\bthm}{\begin{thm}}
\newcommand{\ethm}{\end{thm}}

\newcommand{\bcor}{\begin{cor}}
\newcommand{\ecor}{\end{cor}}
\newcommand{\bprop}{\begin{prop}}
\newcommand{\eprop}{\end{prop}}
\newcommand{\blm}{\begin{lm}}
\newcommand{\elm}{\end{lm}}
\newcommand{\beq}{\begin{equation}}
\newcommand{\eeq}{\end{equation}}
\newcommand{\ber}{\begin{eqnarray}}
\newcommand{\eer}{\end{eqnarray}}

\newcommand{\bproof}{\begin{proof}}
\newcommand{\eproof}{\end{proof}}

\newcommand{\sgn}{\mathop{\mbox{\rm sgn}}}

\newcommand{\bit}{\begin{itemize}}
\newcommand{\eit}{\end{itemize}}
\newcommand{\ben}{\begin{enumerate}}
\newcommand{\een}{\end{enumerate}}
\newcommand{\bdesc}{\begin{description}}
\newcommand{\edesc}{\end{description}}
\newcommand{\beqarrn}{\begin{eqnarray*}}
\newcommand{\eeqarrn}{\end{eqnarray*}}
\newcommand{\bproofof}{\begin{proofof}}
\newcommand{\eproofof}{\end{proofof}}
\newenvironment{rem}{\begin{trivlist}\item[]{\bf
Remark:}\hspace{4mm}}{\end{trivlist}}
\newcommand{\brem}{\begin{rem}}
\newcommand{\erem}{\end{rem}}
\newenvironment{rems}{\begin{trivlist}\item[]{\bf
Remarks}\begin{itemize}}{\end{itemize}\end{trivlist}}
\newcommand{\brems}{\begin{rems}}
\newcommand{\erems}{\end{rems}}
\newtheorem{fact}{Fact}
\newcommand{\bfact}{\begin{fact}}
\newcommand{\efact}{\end{fact}}
\newtheorem{examp}{Example}
\newcommand{\bexamp}{\begin{examp}\rm}
\newcommand{\eexamp}{\end{examp}}
\newtheorem{defn}{Definition}
\newcommand{\bdefn}{\begin{defn}\rm}
\newcommand{\edefn}{\end{defn}}

\newtheorem{alg}{Algorithm}
\newcommand{\balg}{\begin{alg}}
\newcommand{\ealg}{\end{alg}}

\newtheorem{prob}{Problem}
\newcommand{\bprob}{\begin{prob}}
\newcommand{\eprob}{\end{prob}}

\newcommand{\bvtm}{\begin{verbatim}}
\newcommand{\bfig}{\begin{figure}}
\newcommand{\efig}{\end{figure}}
\newcommand{\bcen}{\begin{center}}
\newcommand{\ecen}{\end{center}}

\long\def\comment#1{}

\def \n2{{N_0 \over 2}}

\def \h5{\hspace{0.5in}}

\newcommand{\bs}{{\utwi{s}}}

\newcommand{\bC}{{\utwi{C}}}

\newcommand{\bG}{{\utwi{G}}}
\newcommand{\bH}{{\utwi{H}}}
\newcommand{\bI}{{\utwi{I}}}

%% file: paper_5_8.bbl
\begin{thebibliography}{0}

\bibitem{GA{2006}}
G. A. Shaw, A. M. Siegel, and J. Model, ``Extending the range and performance of non-line-of-sight
ultraviolet communication links," \emph{Defense and Security Symposium. International Society for Optics and Photonics}, Orlando, Florida, USA, Apr. 17, 2006.
\bibitem{ZXu{2008}}
Z. Xu and B. M. Sadler, ``Ultraviolet communications: potential and state-of-the-art," \emph{IEEE Commun. Mag.}, vol. 46, no. 5, pp. 67-73, May 2008.

\bibitem{Michael{2007}}
M. R. Frey, ``Information capacity of the Poisson channel," \emph{IEEE Trans. Inform. Theory}, vol. 37, no. 2, pp. 244-256, Mar. 1991.
\bibitem{Aaron{1988}}
A. D. Wyner, ``Capacity and error exponent for the direct detection photon channel - Part \uppercase\expandafter{\romannumeral1}-\uppercase\expandafter{\romannumeral2}," \emph{IEEE Trans. Inform. Theory}, vol. 34, no. 6, pp. 1449-1471, Nov. 1988.

\bibitem{JCao{2014}}
J. Cao, S. Hranilovic, and J. Chen, ``Capacity-achieving distributions for the discrete-time Poisson channel - Part \uppercase\expandafter{\romannumeral1}-\uppercase\expandafter{\romannumeral2}," \emph{IEEE Trans. Commun.}, vol. 62, no. 1, pp. 194-213, Jan. 2014.
\bibitem{ALapidoth{2009}}
A. Lapidoth and S. M. Moser, ``On the capacity of the discrete-time Poisson channel," \emph{IEEE Trans. Info. Theory}, vol. 55, no. 1, pp. 303-322, Jan. 2009.

\bibitem{SM{2003}}
S. M. Haas and J. H. Shapiro, ``Capacity of wireless optical communications," \emph{IEEE Joural Select. Areas Commun.}, vol. 55, no. 1, pp. 303-322, Jan. 2009.
\bibitem{KChakraborty{2008}}
K. Chakraborty, S Dey, and M. Franceschetti, ``Outage capacity of MIMO Poisson fading channels," \emph{IEEE Trans. Info. Theory}, vol. 54, no. 11, pp. 4887-4907, Nov. 2008.


\bibitem{HDing{2009}}
H. Ding, G. Chen, A. K. Majumdar, B. M Sadler and Z. Xu, ``Modeling of non-line-of-sight ultraviolet scattering channels for communication," \emph{IEEE Journal Select. Areas Commun.}, vol. 27, no. 9, pp. 1535-1544, Dec. 2009.
\bibitem{HZhang{2012}}
H. Zhang, H. Yin, H. Jia, S. Chang, and J. Yang, ``Characteristics of non-line-of-sight polarization ultraaviolet communication channels," \emph{Appl. Opt.}, vol. 51, no. 35, pp. 8836-8372, Dec. 2012.
\bibitem{YSun{2016}}
Y. Sun and Y. Zhan, ``Closed-form impulse response model of non-line-of-sight single-scatter propagation," \emph{J. Opt. Soc. Am. A}, vol. 33, no. 4, pp. 752-757, Apr. 2016.
\bibitem{RDrost{2013}}
R. Drost, T. Moore, and B. Sadler, ``Ultraviolet scattering propagation modeling: analysis of path loss versus range," \emph{J. Opt. Soc. Am. A}, vol. 30, no. 11, pp. 2259-2265, Nov. 2013.
\bibitem{HXiao{2011}}
H. Xiao, Y. Zuo, J. Wu, H. Guo and J. Lin, ``Non-line-of-sight ultraviolet single-scatter propagation model," \emph{Opt. Express}, vol. 19, no. 18, pp. 17864-17875, Aug. 2011.
\bibitem{YZuo{2013}}
Y. Zuo, H. Xiao, J. Wu, Y. Li, and J. Lin, ``Closed-form path loss model of non-line-of-sight ultraviolet single-scatter propagation," \emph{Opt. Lett.}, vol. 38, no. 12, pp. 2116-2118, Jun. 2013.

\bibitem{QHe{2010}}
Q. He, Z. Xu, and B. M. Sadler, ``Performance of non-line-of-signt LED based ultraviolet communication receivers," \emph{Opt. Express}, vol. 18, no. 12, pp. 12 226-12 238, May 2010.

\bibitem{NChatzidiamantis{2010}}
N. Chatzidiamantis, G. K. Karagiannidis and M. Uysal, ``Generailized maximum-likelihood sequence detection for photon-counting free space optical systems," \emph{IEEE Trans. Commun.}, vol. 58, no. 12, pp. 3381-3385, Dec. 2010.
\bibitem{Gong{2015}}
C. Gong and Z. Xu, ``LMMSE SIMO receiver for short-range non-line-of-sight scatterign communication," \emph{IEEE Trans. Wireless Commun.}, vol. 14, no. 10, pp. 5338-5349, Oct. 2015.
\bibitem{ME{2015}}
M. El-Shimy and S. Hranilovic, ``Spatial-diversity imaging receivers for non-line-of-sight solar-blind UV communications," \emph{IEEE/OSA J. Lightw. Technol.}, vol. 33, no. 11, pp. 2246-2255, Jun. 2015.

\bibitem{LLiao{2015}}
L. Liao, Z. Li, T. Lang, and G. Chen, ``UV LED array based NLOS UV turbulence channel modeling and experimental verification," \emph{Opt. Express}, vol. 23, no. 17, pp. 21 825-21 835, Aug. 2015.
\bibitem{Kun{2017}}
K. Wang, C. Gong, D. Zou, and Z. Xu, ``Turbulence channel modeling and non-parametric estimation for optical wireless scattering communication," \emph{IEEE/OSA J. Lightw. Technol.}, vol. 35, no. 13, pp. 2746-2756, Apr. 2017.
\bibitem{GA{2009}}
G. A. Shaw, A. M. Siegel, J. Model, and A. Geboff, ``Deep UV photon-counting detectors and applications," \emph{SPIE Defense, Security, and Sensing. International Society for Optics and Photonics}, Orlando, Florida, USA, Apr. 13, 2009.

\bibitem{GChen{2014}}
G. Chen, L. Liao, Z. Li, R. J. Drost, and B. M. Sadler, ``Experimental and simulated evaluation of long distance NLOS UV communication," \emph{Communicaiton Systems, Networks Digital Signal Processing (CSNDSP), 2014 9th International Symposium on. IEEE}, Manchester, UK, Jul. 23-25, 2014.

\bibitem{AM{2004}}
A. M. Siegel, G. A. Shaw, and J. Model, ``Short-range communication with ultraviolet LEDs," \emph{Optical Science and Technology, the SPIE 49th Annual Meeting International Society for Optics and Photonics}, Denver, Colorado, USA, Aug. 2, 2004.
\bibitem{DH2012}
D. Han, Y. Liu, K. Zhang, P. Luo, and M. Zhang, ``Theoretical and experimental research on diversity reception technology in NLOS UV communication system", \emph{Opt. Express}, vol. 20, no. 14, pp. 15833-15842, Jun. 2012.
\bibitem{MW2014}
M. Wu, D. Han, X. Zhang, F. Zhang, M. Zhang, and G. Yue, ``Experimental research and comparison of LDPC and RS channel coding in ultraviolet communication systems", \emph{Opt. Express}, vol. 22, no. 5, pp. 5422-5430. Feb. 2014.
\bibitem{LG2015}
L. Guo, D. Meng, K. Liu, X. Mu, W. Feng, and D. Han, ``Experimental research on the MRC diversity reception algorithm for UV communication", \emph{Applied Optics}, vol. 54, no. 16, pp. 5050-5056, Jun. 2015.
\bibitem{HQin{2017}}
H. Qin, Y. Zuo, D. Zhang, Y. Li, and J. Wu, ``Received response based heuristic LDPC code for short-range non-line-of-sight ultraviolet communication," \emph{Opt. Express}, vol. 25, no. 5, pp. 5018-5030, Mar. 2017.
\bibitem{KunW{2017}}
K. Wang, C. Gong, D. Zou, X. Jin, and Z. Xu, ``Demonstration of a 400 kbps real-time non-line-of-sight laser-based ultraviolet communication system over 500m," \emph{Chin. Opt. Lett.}, vol. 15, no. 4, pp. 040602, Apr. 2017.

\bibitem{XiaonaLiu{2016}}
X. Liu, C. Gong, S. Li, and Z. Xu, ``Signal characterization and receiver design for visible
light communication under weak illuminance," \emph{IEEE Commun. Lett.}, vol. 20, no. 7, pp. 1349-1352, Jul. 2016.
\bibitem{MY{2004}}
M. Yang, W. E. Ryan, and Y. Li, ``Design of efficiently encodable moderate-length high-rate irregular LDPC codes," \emph{IEEE Trans. Commun.}, vol. 52, no. 4, pp. 564-571, Apr. 2004.
\bibitem{MPCF{2004}}
M. P. C. Fossorier, ``Quasi-Cyclic low-Decsity parity-check codes from circulant permutation matrices," \emph{IEEE Trans. Inform. Theory}, vol. 50, no. 8, pp. 1788-1793, Aug. 2004.
\bibitem{JC{2005}}
J. Chen, A. Dholakia, E. Eleftheriou, M. P. C. Fossorier, and X. Y. Hu, ``Reduced-complexity decoding of LDPC codes," \emph{IEEE Trans. Commun.}, vol. 53, no. 8, pp. 1288-1299, Aug. 2005.

\end{thebibliography}
